\input harvmac

\noblackbox
%\draftmode
%

\def\IR{\relax{\rm I\kern-.18em R}}

%%%
\def\abstract#1{
\vskip .5in\vfil\centerline
{\bf Abstract}\penalty1000
{{\smallskip\ifx\answ\bigans\leftskip 2pc \rightskip 2pc
\else\leftskip 5pc \rightskip 5pc\fi
\noindent\abstractfont \baselineskip=12pt
{#1} \smallskip}}
\penalty-1000}  
\def\us#1{\underline{#1}}
\def\hth/#1#2#3#4#5#6#7{{\tt hep-th/#1#2#3#4#5#6#7}}
\def\nup#1({Nucl.\ Phys.\ $\us {B#1}$\ (}
\def\plt#1({Phys.\ Lett.\ $\us  {B#1}$\ (}
\def\cmp#1({Commun.\ Math.\ Phys.\ $\us  {#1}$\ (}
\def\prp#1({Phys.\ Rep.\ $\us  {#1}$\ (}
\def\prl#1({Phys.\ Rev.\ Lett.\ $\us  {#1}$\ (}
\def\prv#1({Phys.\ Rev.\ $\us  {#1}$\ (}
\def\mpl#1({Mod.\ Phys.\ Let.\ $\us  {A#1}$\ (}
\def\ijmp#1({Int.\ J.\ Mod.\ Phys.\ $\us{A#1}$\ (}
%%%

\def\cR{{\cal R}}

\def\ao{a^{(0)}}
\def\bo{b^{(0)}}
\def\co{c^{(0)}}
\def\a1{a^{(1)}}
\def\b1{b^{(1)}}
\def\c1{c^{(1)}}

%%%%%%%%%%%%%%%%%%%%%%%%%%%%%%%%%%%%%%%%%%%%%
\vskip-2cm
\Title{\vbox{
\rightline{\vbox{\baselineskip12pt\hbox{LMU-TPW-98-14}
                                   \hbox{hep-th/9808126}}}}}
{$AdS_5\times S^5$ Black Hole Metric at ${\cal O}$($\alpha'^3)$}
\vskip 0.3cm
\centerline{Jacek Pawe{\l}czyk$^{a,b,*}$ and Stefan Theisen$^{b}$}
\vskip 0.6cm
\centerline{$^a$ \it Institute for Theoretical Physics, Warsaw University}
\vskip-.2cm
\centerline{\it Hoza 69, PL-00-681 Warsaw, Poland}
\vskip.4cm
\centerline{$^b$ \it Sektion Physik, Universit\"at M\"unchen, 
D-80333 Munich, Germany}
\vskip 0.3cm
\abstract{We compute ${\cal O}(\alpha'^3)$ corrections to the 
$AdS_5\times S^5$ black hole metric.
We find that the radius of the $S^5$ depends on the radial $AdS_5$
coordinate. This completes the computation of
Gubser, Klebanov and Tseytlin (hep-th/9805156). The fact that the 
metric no longer factorizes should modify
the value of the Wilson line at finite temperature 
and the glueball mass spectrum.
}
\vskip5cm
\noindent\hrule
\vskip.1cm
\noindent $^*$ Supported by the Alexander-von-Humboldt Foundation.

\vfill\eject

\parskip=4pt plus 15pt minus 1pt
\baselineskip=15pt plus 2pt minus 1pt

Sparked by Maldacena's conjecture \ref\Maldacena{J. Maldacena, 
The large N limit of superconformal field theories and supergravity, 
hep-th/9711200}
there has recently been a resurgence of interest in supergravity in
anti-de-Sitter space. In the simplest case the type IIB vacuum is the 
direct product $AdS_5\otimes S^5$ \ref\Schwarz{J. Schwarz, Covariant
field equations of chiral $N=2$, $D=10$ supergravity, 
Nucl. Phys. B226 (1983) 269}.
Many of the subsequent papers on various 
aspects of Maldacena's conjecture were based on the 
leading order supergravity actions. 
However, since the conjecture
refers to the complete string theory, one should 
consider the string corrections 
to the 10D supergravity action. The first corrections occur at order 
$(\alpha')^3$ and have been known for a long time 
\ref\othree{M.T.~Grisaru and D.~Zanon, Sigma-model superstring 
corrections to the Einstein-Hilbert action, Phys. Lett B177 (1996) 347; 
M.D.~Freeman, C.N.~Pope, M.F.~Sohnius and K.S.~Stelle, Higher-order
sigma-model counterterms and the effective action for superstrings, 
Phys. Lett. B178 (1996) 199; 
D.J.~Gross and E.~Witten, Superstring modifications of
Einstein equations, Nucl. Phys. B277 (1986) 1}. 
Taking theses effects into 
account, as shown by Banks and Green 
\ref\BG{T.~Banks and M.~Green,
On perturbative effects in $AdS_5\times S^5$ string theory and 
$d=4$ SUSY Yang-Mills,
JHEP 05 (1998) 002, hep-th/9804170},
does not change the metric in the extremal $AdS_5\otimes S^5$ case. 
This was subsequently verified to all orders in $\alpha'$ in 
\ref\KR{R.~Kallosh and A.~Rajaraman,
Vacua of M theory and string theory, hep-th/9805041}.
In the non-extremal case this is however no longer true and one 
is faced with the task to compute the corrections to the metric and 
the other background fields, such as the dilaton and the 
anti-symmetric tensor field. This problem was addressed recently
by Gubser, Klebanov and Tseytlin 
\ref\GKT{S.~Gubser, I.~Klebanov and A.~Tseytlin, Coupling constant 
dependence in the thermodynamics of N=4 supersymmetric Yang-Mills
theory, hep-th/9805156; see in particular the note added in 
the revised version.}. Their analysis, which was restricted
to the $AdS_5$ part of the metric, turns out 
to be sufficient for the computation of the corrections of 
the free energy. However, the corrections to the full ten-dimensional
metric has not been found in \GKT, as has been erroneously assumed 
in several subsequent papers. Specifically, the dynamics of the conformal 
factor was not found. 
Here we reconsider the issue for the full ten-dimensional metric and 
show that at ${\cal O}(\alpha'^3)$ it no longer factorizes. 

The starting point for the analysis is
the low-energy supergravity action in the Einstein frame
\eqn\actionfour{
S={N^2\over 16\pi^7}\int d^{10}x\sqrt{-g}
\Bigl\lbrace\cR-{1\over2}(\partial\phi)^2
+\gamma e^{-{3\over2}\phi}W
-{1\over 4\cdot 5!}{1\over N^2}F_5^2\Bigr\rbrace
}
where we have defined
\eqn\tgamma{
\gamma\equiv {1\over 8}\zeta(3)(g_s N)^{-3/2}\,.
}
In the Maldacena limit $(g_s N)^{1/2}\sim\alpha'$.
Note that the normalization is such that $F_5\sim N$. For details, 
in particular for a discussion on the form $W\sim{\cal C}^4$ 
(${\cal C}$ is the Weyl tensor) of the 
eighth derivative term and the subtleties with the self-duality 
of $F_5$ we refer to \GKT\ and references quoted therein.

We make the most general ansatz 
compatible with the symmetries of the problem:
\eqn\ouransatz{
ds^2=H^2(r)\Bigl(K^2(r)d\tau^2+P^2(r)dr^2+\sum_{i=1}^3 dx_i^2\Bigr)
+L^2(r)d\Omega_5^2\,.
}
We first show that it is not possible to keep the radius of 
the $S^5$ fixed to 1 (if $H$ is fixed), i.e. $L(r)=1$ is not
a solution to the equations of motion. 
We will then find the correct solution of the 
equations of motion following from the ansatz \ouransatz.
As the authors of \BG\ and \GKT\ we assume that the 
vielbein components of $F_5$ do not change.

After rescaling $ds^2\to\Lambda^2 ds^2$ the 
part of the action containing $\Lambda$ is 
\eqn\Lambdapart{
\eqalign{
S&\supset \int d^{10}x\sqrt{g}\Lambda^{10}\left\lbrace 
\Lambda^{-2}R-18\,\Lambda^{-3}\nabla^2\Lambda-54\,\Lambda^{-4}
(\nabla\Lambda)^2
+\gamma \Lambda^{-8}W\right\rbrace\cr
&=\int d^{10}x\sqrt{g}\left\lbrace\Lambda^8\cR+72\,\Lambda^6(\nabla\Lambda)^2
+\gamma \Lambda^2 W\right\rbrace\,.}
}
Here $\cR$ is the curvature scalar of the metric 
$ds^2=ds_1^2\oplus d\Omega_5^2$, i.e. the metric \ouransatz\ with
$L(r)=1$.
We have neglected terms ${\cal O}(\gamma^2)$. They will not enter
the argument.
Due to the direct sum structure we have 
$\cR=\cR_1+\cR_2=\cR_1+20$, the latter part coming from $S^5$ 
($\cR_{S^n}=n(n-1)$). 

Consider now the equation of motion for $\Lambda$,
\eqn\eqOmega{
8\,\Lambda^7(\cR_1+20)-6\cdot 72\,\Lambda^5(\nabla\Lambda)^2
-2\cdot 72\,\Lambda^6\nabla^2\Lambda+2\,\gamma\Lambda W = 0\,.
}
The solution for $\Lambda$ will be of the form
\eqn\ansatzOmega{
\Lambda=\Lambda^{(0)}+\gamma\Lambda^{(1)}+\dots
}
where $\Lambda^{(0)}$ and $\Lambda^{(1)}$ are both ${\cal O}(\alpha'^0)$. 
We likewise expand
\eqn\ansatzR{
\cR_1=\cR_1^{(0)}+\gamma\cR_1^{(1)}+\dots
}
and
\eqn\ansatzR{
W=W^{(0)}+\gamma W^{(1)}+\dots
}
If $\Lambda\equiv 1$ is a solution,   
the equation of motion for $\Lambda$ will be satisfied if
\eqn\eqR{
\cR_1^{(0)}+20=0 \qquad{\rm and}\qquad
4\cR_1^{(1)}+W^{(0)}=0
}
hold. In fact $\cR_1^{(0)}=-20$ but the latter equation is not satisfied, because 
\eqn\Wzero{
W^{(0)}=180{r_0^{16}\over r^{16}}\qquad{\rm and}\qquad
\cR_1^{(1)}=180{r_0^{16}\over r^{16}}
}
for the metric given in \GKT. This completes the proof that the
ten-dimensional metric is not of the form $ds_1^2\oplus d\Omega_5^2$.

Next we will solve the equations of motion
following from the general ansatz \ouransatz\ which shows explicitly that 
a direct product geometry is not a solution. 
For ease of comparison with \GKT\ we choose the following parameterization
for the functions $H(r),K(r),P(r),L(r)$:
\eqn\expform{
\eqalign{
H(r)&=r\,,\cr
K(r)&=e^{a(r)+4 b(r)}\,,\cr
P(r)&=e^{b(r)}\,,\cr
L(r)&=e^{c(r)}\,.}}
In terms of these functions the lowest, i.e. zeroth order in $\alpha'$,
contribution to the action is ($'=\partial_r$)
\eqn\action{
\eqalign{
S=\int dr\Bigl\lbrace & 4 r^5 \Bigl(5-2 e^{-8 c}\Bigr)e^{a+5b+3c}\cr
&+\Bigl(-2r(2+r a')+10 r^3 c'(a'+4 b'+2 c')\Bigr)e^{a+3b+5c}\cr
&\qquad -2\Bigl(r^3(a'+4b'+5c')e^{a+3b+5c}\Bigr)'\Bigr\rbrace.
}}
We have dropped an overall factor ${N^2\over\pi^7}{\rm Vol}(S^5)
{\rm Vol}({\bf R}^{3,1})={N^2\over\pi^4}{\rm Vol}({\bf R}^{3,1})$.
The expression for the $W$ term is too long to reproduce here. 

Since we can consistently find solutions to 
${\cal O}(\gamma)$ only, we write 
\eqn\expansiona{
a(r)=\ao(r)+\gamma\a1(r)
}
and likewise for $b(r)$ and $c(r)$, suppressing higher order terms in $\gamma$. 
Perturbation in $\gamma$ requires the zeroth order solutions.
They are
\eqn\abczeroth{
\eqalign{
\ao(r)&=-\log(r^2)+{5\over2}\log(r^4-r_0^4)\cr
\bo(r)&=-{1\over2}\log(r^4-r_0^4)\cr
\co(r)&=0}
}
The equations of motion for the first order (in $\gamma$)
corrections get contributions from the term 
$\propto\gamma W$ in the action eq.\actionfour. They are, up
to a factor $\gamma$, 
\eqn\contW{
540(19 r_0^4-16 r^4){r_0^{12}\over r^{13}}\,,\qquad 
540(79 r_0^4-64 r^4){r_0^{12}\over r^{13}}\,,\qquad
900{r_0^{16}\over r^{13}}}
for the equation for $a(r)$, $b(r)$ and $c(r)$, respectively.
The equations can now be easily solved with the ansatz
\eqn\ansatza{
\a1(r)=a_0+a_1{r_0^4\over r^4}+a_2{r_0^8\over r^8}+a_3{r_0^{12}\over r^{12}}+\dots
}
and likewise for $\b1(r)$ and $\c1(r)$. 
It turns out that higher powers in ${r_0\over r}$ beyond the 
ones displayed will not contribute.
The results are
\eqn\solutions{
\eqalign{
\a1(r)&=-{1625\over 8}{r_0^4\over r^4}
-175{r_0^8\over r^8}+{10005\over 16}{r_0^{12}\over r^{12}}\cr
\b1(r)&={325\over 8}{r_0^4\over r^4}+{1075\over 32} {r_0^8\over r^8}
-{4835\over 32}{r_0^{12}\over r^{12}}\cr
\c1(r)&={15\over 32}{r_0^8\over r^8}(1+{r_0^4\over r^4})}
}
$a_0$, which is undetermined and related to rescaling of time, 
has been set to zero \GKT.
The equation for the first correction of the dilaton 
$\phi=-\log(g_s)+\gamma\phi^{(1)}+\dots$ is the same as in \GKT\ and 
leads to
\eqn\dilaton{
\phi^{(1)}(r)=-{45\over 8}\Bigl({r_0^4\over r^4}+{1\over 2}{r_0^8\over r^8}
+{1\over 3}{r_0^{12}\over r^{12}}\Bigr)}

We can also give the necessary reparameterization that 
transforms the five-dimensional $AdS_5$ part of the metric 
as computed here to the one computed in \GKT:
\eqn\repara{
r\to r\Bigl[1-\gamma{25\over 32}\Bigl({r_0^8\over r^8}
+{r_0^{12}\over r^{12}}\Bigr)\Bigr]\,,
\qquad r_0\to r_0\Bigl(1-{25\over 16}\gamma\Bigr)
}
The resulting metric is
\eqn\remetric{
ds^2=e^{-10/3\nu(r)}H^2(r)\Bigl(K^2(r)d\tau^2+P^2(r)dr^2+
\sum_{i=1}^3 dx_i^2\Bigr)
+e^{2\nu(r)}d\Omega_5^2}
where $H(r)=r$, $K(r)$ and $P(r)$ are as in \GKT\ and to ${\cal O}(\gamma)$
\eqn\resnu{
\nu(r)=\gamma{15\over 32}{r_0^8\over r^8}(1+{r_0^4\over r^4})}

There are several applications of our result. First of all, we have
reconsidered the corrections of thermodynamic quantities, 
following ref.\GKT. It turns out that the correction to the 
free energy does not change. 
For comparison we give some results. For the temperature we find
\eqn\temperature{
2\pi T=2r_0\Bigl(1+{265\over 16}\gamma\Bigr)
}
The action is
\eqn\actionI{
I={N^2\over 4\pi^2}\beta {\rm Vol}({\bf R}^3)(r_{\rm max}^4-r_0^4)
\left(1-{325\over 4}\gamma
\left[{r_0^4\over r_{\rm max}^4}
+{\cal O}({r_0^8\over r_{\rm max}^8})\right]\right)
}
For the free energy we find
\eqn\freenergy{
F=-{\pi^2\over 8}N^2 V_3 T^4(1+15\gamma)
}
which agrees with the result in \GKT. 
For an independent argument why the 
value of the free energy does not change, see the note added in \GKT.

One can also see that the scalar glueball spectrum 
(without KK modes) \ref\ooguri{C.~Csaki, H.~Ooguri, Y.~Oz and J.~Terning,
Glueball mass spectrum from supergravity, hep-th/9806021; 
H.~Ooguri, H.~Robins and J.~Tannenhauser, 
Glueballs and their Kaluza-Klein cousins, hep-th/9806171} is unchanged. 
The reason is that inclusion of the $L^2$ factor in \ouransatz\ 
does not influence the relevant equations of motions. 
However there will be corrections to the other glue-ball masses and to KK
glueballs. 
Likewise, the coefficients of the ${\cal O}(\gamma)$
corrections to the Wilson loop 
at finite temperature \ref\DO{J.~Greensite and P.~Olesen, 
Remarks on the heavy quark potential in the supergravity
approach, hep-th/9806235; 
H.~Dorn and H.J.~Otto, 
$q\bar q$ potential from AdS-CFT relation at $T\geq 0$:
dependence on orientation in space and higher curvature corrections, 
hep-th/9807093} will change.

\vskip1cm

\noindent
{\bf Acknowledgment:} We want to thank Arkady Tseytlin for useful 
correspondence.
S.T. also thanks D. Maison for a clarifying discussion. 
Our work is supported by GIF, the German-Israeli Foundation
for Scientific Research and by the European Commission TMR programme
ERBFMRX-CT96-0045.

\listrefs
\end

\bye